\documentclass[conference]{IEEEtran}
\IEEEoverridecommandlockouts
\usepackage{cite}
\usepackage{amsmath,amssymb,amsfonts}
\usepackage{algorithmic}
\usepackage{graphicx}
\usepackage{textcomp}
\usepackage{xcolor}
\usepackage{tabularx}
\usepackage{datatool}
\usepackage{float}

\def\BibTeX{{\rm B\kern-.05em{\sc i\kern-.025em b}\kern-.08em
    T\kern-.1667em\lower.7ex\hbox{E}\kern-.125emX}}

\begin{document}


\thispagestyle{empty}

\begin{huge}
IEEE Copyright Notice
\end{huge}

\vspace{5mm} 

\vspace{5mm} 

\begin{large}
© 2024 IEEE.  Personal use of this material is permitted.  Permission from IEEE must be obtained for all other uses, in any current or future media, including reprinting/republishing this material for advertising or promotional purposes, creating new collective works, for resale or redistribution to servers or lists, or reuse of any copyrighted component of this work in other works.
\end{large}

\vspace{5mm} 

\begin{large}
\textbf{Submitted to:} IEEE CSCE 2024 · July 22-25, 2024 -  Las Vegas, NV
https://ieeesoutheastcon.org/ 
\end{large}

\vspace{5mm} 
Preprint Version, April 04, 2024

\vspace{5mm} 

\newcolumntype{L}[1]{>{\raggedright\arraybackslash}p{#1}}
\newcolumntype{C}[1]{>{\centering\arraybackslash}p{#1}}
\newcolumntype{R}[1]{>{\raggedleft\arraybackslash}p{#1}}

\clearpage
\pagenumbering{arabic}

\title{The Shifting Landscape of Cybersecurity: The Impact of Remote Work and COVID-19 on Data Breach Trends}

\makeatletter
\newcommand{\linebreakand}{
  \end{@IEEEauthorhalign}
  \hfill\mbox{}\par
  \mbox{}\hfill\begin{@IEEEauthorhalign}
}


\makeatother
\author{

\IEEEauthorblockN{Murat Ozer}
\IEEEauthorblockA{\textit{School of Information Technology} \\
\textit{University of Cincinnati}\\
Cincinnati, Ohio, USA \\
m.ozer@uc.edu}
\and
  \IEEEauthorblockN{Yasin Kose}
  \IEEEauthorblockA{\textit{Friedrich-Alexander-Universität } \\
    \textit{Cybercrime and Forensic Computing}\\
    Erlangen, Germany \\
    yasin.koese@fau.de}
\linebreakand 
  \IEEEauthorblockN{Mehmet Bastug}
  \IEEEauthorblockA{\textit{CJ and Criminology Department} \\
    \textit{University of Scranton}\\
    Scranton, PA, USA \\
    mehmet.bastug@scranton.edu}
\and 
\IEEEauthorblockN{Goksel Kucukkaya}
\IEEEauthorblockA{\textit{School of Information Technology} \\
\textit{University of Cincinnati}\\
Cincinnati, Ohio, USA \\
kucukkgl@ucmail.uc.edu}
\and 
\IEEEauthorblockN{Eva Ruhsar Varlioglu}
\IEEEauthorblockA{\textit{School of Criminal Justice} \\
\textit{University of Cincinnati}\\
Cincinnati, Ohio, USA \\
varliorr@mail.uc.edu}
}

\IEEEoverridecommandlockouts
\IEEEpubid{\makebox[\columnwidth]{
\textit{Preprint Version} ~\copyright\textit{2024 IEEE} \hfill} \hspace{\columnsep}\makebox[\columnwidth]{ }}

\maketitle

\thispagestyle{plain}
\pagestyle{plain}

\begin{abstract}

This study examines the impact of the COVID-19 pandemic on cybersecurity and data breaches, with a specific focus on the shift toward remote work. The study identifies trends and offers insights into cybersecurity incidents by analyzing data breaches two years before and two years after the start of remote work. Data was collected from the Montana Department of Justice Data Breach database and consisted of data breaches that occurred between April 2018 and April 2022. The findings inform best practices for cybersecurity preparedness in remote work environments, aiding organizations to enhance their defenses. Although the study's data is limited to Montana, it offers valuable insights for cybersecurity professionals worldwide. As remote work continues to evolve, organizations must remain adaptable and vigilant in their cybersecurity strategies.
\end{abstract}
 
\section{Introduction}
Starting in early 2020, the world faced an unprecedented global health crisis that disrupted industries and impacted lives worldwide \cite{weil2020it}. To control the spread of the crisis and "flatten the curve," organizations and businesses swiftly shifted to remote work, making it the new norm virtually overnight. This sudden reliance on technology for business continuity and remote interaction highlighted the crucial role of connectivity in maintaining operations. However, this shift also brought inherent security challenges as employees started conducting business outside traditional office environments and on potentially untrusted devices. With only a fraction of businesses having a cybersecurity policy in place \cite{pranggono2021covid}, the increased risk of cyberattacks during this period was significant \cite{deo2020covid}.
The shift to remote work, commonly known as "working from home" (WFH), posed a new challenge for IT security professionals \cite{bispham2021cybersecurity}. Examining data breach trends before and during the global pandemic can provide insights to understand this new challenge and better prepare our cyber security practices/responses based on these data-driven policies. Such an approach can also ensure resilience across multiple platforms in the face of evolving cybersecurity threats. Thus, closely examining data breach information, patterns, and commonalities can benefit organizations and their users, enabling improved preparedness for potential future incidents.
Given this context, this study aims to bridge the knowledge gap and provide a broader understanding of the impact of data breaches in a four-year period. The World Health Organization (WHO) declared the coronavirus outbreak a pandemic on March 11, 2020. Soon after the declaration, organizations from all over the world began switching to remote work. We used the beginning of April 2020 as the milestone for remote work. We collected the data two years before and two years after that milestone. We compared the two time periods to understand how switching to remote work impacted data breach trends. The insights gained will contribute to a better understanding of best practices and strategies to enhance cybersecurity preparedness. As such, the findings of this study will be valuable in shaping policies, procedures, and network controls for IT security professionals in the event of a similar magnitude crisis in the future.

\section{Relevant Studies}
The rapid adoption of remote work in response to the COVID-19 pandemic has significantly changed how organizations operate. While remote work offers numerous benefits, such as increased flexibility and productivity, it also introduces new challenges, particularly in terms of maintaining information security and protecting against data breaches. Many scholars also studied the adverse effects of remote work during COVID-19. Therefore, we first conducted a brief literature review to provide an overview of relevant studies and research on the impact of remote work on cybersecurity, highlighting key findings, challenges, and best practices. 

\subsection{\textit{The impacts of remote work}}

A number of studies have investigated the impact of remote work on cybersecurity incidents. For instance, Škiljić (2020) explored cybersecurity risks associated with remote work and emphasized the importance of adapting cybersecurity controls and policies such as implementing multi-factor authentication, enhancing security in applications while avoiding insecure collaboration tools, deploying encrypted VPN solutions, conducting regular security audits, enforcing robust remote access controls, and prioritizing employee cybersecurity education to mitigate these risks \cite{skiljic2020cybersecurity}. By taking into account the technical side of remote work, Rakha (2023) examined the challenges associated with securing remote access to organizational resources \cite{rakha2023ensuring}. The author discussed the importance of implementing multi-factor authentication, virtual private networks (VPNs), and secure remote desktop protocols to establish secure connections between remote workers and the corporate network \cite{rakha2023ensuring}. 
Dangheralou and Jahankhani (2022) analyzed the impact of remote work on the effectiveness of security controls \cite{dangheralou2021impact}. They found that certain controls, such as network perimeter defenses, might be less effective in a remote work environment. The authors suggested the adoption of cloud-based security solutions and endpoint protection tools to address these limitations \cite{dangheralou2021impact}. On the other hand, by focusing on human error in cyber security, Sidor-Rzaqdkowska (2022) investigated the role of employee behavior in remote work \cite{sidor2022human}. They emphasized the need for strong security awareness programs, regular training, and clear security policies to foster a culture of cybersecurity among remote workers and reduce the risk of human errors leading to data breaches \cite{sidor2022human}.
Borkovich and Skovira (2020) focused on the cybersecurity challenges that arise in the context of remote work environments \cite{borkovich2020working}. They examined the unique security issues associated with remote work arrangements and offered insights into how organizations can address these challenges. Their research sheds light on the evolving landscape of cybersecurity in the remote work setting. Likewise, Nwankpa and Datta (2023) conducted a study that focused on the relationship between remote work and cybersecurity \cite{nwankpa2023remote}. The research specifically examined employee awareness and practices regarding cybersecurity in the context of remote work. By evaluating employee awareness and behaviors related to cybersecurity, the study aimed to shed light on the potential vulnerabilities and risks associated with remote work and identify areas where organizations can improve employee training and awareness to enhance their cybersecurity defenses. 
Ahmad (2020) emphasized the role of cybersecurity policies and practices in mitigating data breach risks by stressing the importance of developing and implementing proactive security strategies, effective computer security plans, the integration of cloud-based systems, and the use of encryption to safeguard organizational assets and resources in the face of evolving cyber threats (Ahmad, 2020). Finally, Nafea and Almaiah (2021) comprehensive literature offers a review of the topic of security risks and countermeasures in remote work. Their study focused on understanding the security challenges associated with remote work, provided an overview of various security risks remote workers face, and explored countermeasures to mitigate these risks, offering a valuable resource for addressing security concerns in remote work environments \cite{alnafea2021cyber}.

\subsection{\textit{Increased level of cyberattacks}}
Certain scholars examined the impact of remote work on the frequency of cyberattacks. Within this framework, Al-Qahtani and Cresci (2022) conducted a study on the impact of the COVID-19 pandemic on cyberattacks \cite{alqahtani2022covid}. They found a significant increase in phishing attacks and social engineering scams targeting remote workers, highlighting the need for robust security awareness training and regular updates on emerging threats. 
By analyzing reported incidents Jawaid (2022) explored the relationship between remote work and data breaches. Their study answers questions about whether remote work impacts data security and privacy \cite{jawaid2022increase}. They reported data breach incidents and provided insights into the factors contributing to these breaches in the context of remote work. Ansback and Sharton (2020) investigated the influence of remote work on insider threats. Their findings revealed that remote employees might exhibit different behavioral patterns that could increase the risk of insider attacks \cite{ansbach2020preventing}. They emphasized the need for continuous monitoring and user behavior analytics to detect and prevent such incidents. Similarly, Homoliak et al. (2019) conducted a case study analysis to examine the relationship between remote work and insider threats. The research helps to understand specific cases about how remote work scenarios can contribute to insider threats within organizations \cite{homoliak2020insight}. By analyzing real-world instances, their study provides valuable insights into the challenges and security considerations associated with remote work. 

\subsection{\textit{The effectiveness of current security implications}}
 
Other studies also discussed the risks associated with remote work and the possible technical safeguards against them. For example, Rah (2023) examined the relationship between remote work and adopting Bring Your Own Device (BYOD) policies (Rah, 2023). This study highlighted antivirus tools, mobile device management systems, virtual private networks, firewalls, secure Wi-Fi, device encryption, regular patching, and similar security controls to mitigate the potential risks associated with personal devices used for work purposes. Similarly, Sood et al. (2018) comprehensively analyzed remote desktop protocol (RDP) attacks \cite{sood2018evidential}. They emphasized the significance of securing RDP connections through strong passwords, network segmentation, and two-factor authentication to prevent unauthorized access and potential data breaches. Likewise,  Ramadan et al. (2021) investigated the impact of remote work on the effectiveness of intrusion detection systems (IDS). Their findings indicated the need for fine-tuning IDS algorithms and rules to account for the changes in network traffic patterns and the dynamic nature of remote work environments \cite{ramadan2021cybersecurity}. 
Larsson and Qollakaj (2023) investigated the impact of the shift to remote work and the usage of VPNs \cite{larsson2023cybersecurity}. Their findings indicated that the surge in remote work has led to a rise in VPN attacks. These attacks often exploit vulnerabilities in VPN systems, so companies are advised to patch their systems. Advanced Persistent Threat (APT) groups have taken advantage of these vulnerabilities, establishing persistent and stealthy access to networks used by remote workers via VPNs. To mitigate these risks and fortify VPN systems and private networks, countermeasures like implementing enforced Multi-Factor Authentication (MFA) and adding multiple layers of defense are recommended. 
Finally, Ngee(2022) conducted a comparative analysis focusing on cybersecurity measures in the remote work environment. The study compared various cybersecurity practices and measures across remote work settings to assess their effectiveness. By conducting a comparative analysis, the authors provided insights into the best practices for implementing cybersecurity measures in remote work scenarios, helping organizations enhance their security posture in this evolving work landscape \cite{ngere2022cybersecurity}. 

\subsection{\textit{Summary of the literature review}}
Previous research can be categorized into two primary domains: (1) security threats associated with remote employees and (2) those related to the technologies used while working remotely. In particular, the studies reviewed collectively emphasize the unique challenges faced by organizations as they transition to remote work environments and the need for robust security measures to mitigate risks. As a result, the findings indicate that remote work introduces vulnerabilities related to home networks, personal devices, and human behavior, which malicious actors often exploit through phishing attacks, social engineering scams, and insider threats. Therefore, organizations should focus on implementing strong authentication mechanisms, encryption protocols, secure remote access solutions, and robust security awareness programs to address these challenges. Additionally, measures such as network segmentation, intrusion detection systems, and monitoring user behavior can play crucial roles in maintaining information security in remote work settings.

\subsection{\textit{Current Study}}
This study aims to investigate the impact of remote work on cybersecurity within the context of the pandemic. The sudden shift to remote work has become the new norm for many organizations and individuals. In addition, the reliance on technology and online services has increased significantly. This transition has raised concerns about potential vulnerabilities and security risks associated with remote work. The current study seeks to understand the trends in data breaches that occurred before and during the pandemic. By examining the trends and patterns in data breaches, the study aims to provide insights into best practices for cybersecurity preparedness in remote work environments. The findings of this study will contribute to the development of effective strategies and policies to mitigate cybersecurity risks and ensure the resilience of organizations in the face of future challenges.

\section{Methodology}
This study sourced its data from the Montana Department of Justice (DOJ) Data Breach database, which is responsible for documenting and monitoring data breaches affecting Montana residents. The database offers comprehensive insights into various aspects of these breaches. The DOJ Montana Data Breach database is a well-established resource in cybersecurity and data breach research \cite{montanajustice}, boasting a rich repository of data spanning several years. This makes it an invaluable tool for studying data breach trends.
The dataset we used in this study contains 2,140 recorded data breach incidents between April 1, 2018, and April 1, 2022. We considered April 1, 2020, as the starting date of the remote work period, as the World Health Organization declared the outbreak as a pandemic on March 27, 2020. Many organizations switched to remote work soon after the declaration. We collected data covering two years before and two years after the start of the remote work. We removed the cases from the dataset when information about any studied variables was missing. 
The dataset contains information about the business name, start and end date of each breach, the date it was reported, the number of Montanans affected, and a link to the notification letter sent to customers. The notification letters involve brief information about the incident and provide customers with some resources they can use to protect themselves from identity fraud. Details about the incidents were not coded in the database. Some letters provided detailed information regarding the attack vectors, while others provided only superficial information, making the letters not suitable for coding. We randomly selected 100 letters from each year and analyzed them to identify what type of services were provided to clients by the companies subjected to data breaches. 

\subsection{Analysis and Results}
Table 1 provides a broad perspective on the pandemic's influence on the frequency of reported data breaches. The chart illustrates a sharp rise in data breaches during the pandemic. The number of incidents after the outbreak doubled compared to the period before. Breaches range in size with varying levels of complexity and service interruption. The analysis shows that the data of 242,973 customers were compromised in data breaches two years before the pandemic. The number increased to 403,302 in the two-year period during the pandemic. There is a huge discrepancy in the number of people affected by the data breach incidents. In 67\% / of all incidents within the four-year period, less than 10 people were affected. There are two incidents in which more than a hundred thousand client’s data were exposed. The median value is 3 in the first study period and 4 in the second period.

\begin{table*}[ht]
    \centering
    \begin{tabular}{lccc}
        \hline
        \textbf{Category} & \textbf{Total (N)} & \textbf{Before remote work (N)} & \textbf{After remote work (N)} \\
        \hline
        Total number of incidents & 2127 & 686 & 1441 \\
        Total number of people affected & 646275 & 242973 & 403302 \\
        Median number of people affected & 4 & 3 & 4 \\
        Average number of people affected & 304 & 354 & 280 \\
        Median incident duration (days) & 5 & 9 & 4 \\
        Average incident duration (days) & 35 & 44 & 30 \\
        Median reporting time (days) & 123 & 106 & 131 \\
        Average reporting time (days) & 152 & 132 & 162 \\
        \hline
    \end{tabular}
    \caption{Summary of incidents and impact}
    \label{tab:incident_summary}
\end{table*}

There is a remarkable change in the incident durations between the two periods. Although the number of incidents more than doubled in the second period, the median incident duration decreased from nine to four days. The average duration also follows a similar trend, decreasing from 44 to 30. These findings show that cybersecurity professionals have become more vigilant against detecting data breaches. Increased cyberattacks might have been associated with increased detection and response capabilities. Another explanation could be that attackers achieve their objectives relatively more quickly. Our analysis also shows that it took longer for organizations to notify their customers about the incidents. The median reporting time was increased from 106 to 131 in the second period compared to the first. This could be a result of decreased workflow efficiency due to substantial changes in work arrangements and modalities.

\begin{table*}[ht]
    \centering
    \begin{tabular}{lccc}
        \hline
        \textbf{Year} & \textbf{Total Employed Cybersecurity Workforce} & \textbf{Total Cybersecurity Job Openings} & \textbf{Supply/Demand Ratio} \\
        \hline
        2018 & 940,653 & 538,438 & 95\% \\
        2019 & 974,192 & 560,239 & 93\% \\
        2020 & 1,003,958 & 518,148 & 90\% \\
        2021 & 1,000,702 & 582,818 & 82\% \\
        2022 & 1,102,311 & 720,727 & 70\% \\
        \hline
    \end{tabular}
    \caption{Cybersecurity Supply and Demand over the Years}
    \label{tab:cybersecurity_supply_demand}
\end{table*}

The statistics on the cybersecurity workforce between 2018 and 2022 are presented in Table 2. The table indicates that the supply/demand ratio has been steadily decreasing over the years, while the number of employed individuals in cybersecurity-related positions has slightly increased each year. The data clearly shows that there is a higher demand for cybersecurity workers than the available supply. In 2022, there were only enough cybersecurity professionals to fill 70\% of the demand in both the public and private sectors in the US. It is worth noting that there is an interesting trend regarding the overall count of available cybersecurity job positions. In fact, the total number of job openings in this field for each year is more than half of the number of people employed in the workforce during the same year. This data, coupled with the decreased supply/demand ratio, indicates that organizations may be struggling to either find qualified candidates or retain them within the organization.
After reviewing the notification documents, some emerging trends were evident. Over 66\% of the breaches offered some form of Identity/Credit Monitoring for the victim(s) in both datasets. The major credit monitoring companies, Experian, Equifax, TransUnion, and Kroll, benefited the most from these breaches with the added business. One compelling similarity between the data sets was the employment of an outside forensics firm or third-party security professionals to aid in the breach discovery, details, and planned steps to remedy the intrusion. Some of the organizations were very large, and the breach affected hundreds, if not thousands of people. Though around 26\% of organizations did not exercise the option of seeking outside counsel, there was no mention of an ‘in-house’ forensics team. The notification letters sent to customers seem constructed from a general language template. Each breach was different, but the notification letters' wording, structure, and delivery show similarities. Most letters were vague in the description of the incident. For example, one breach notification letter only stated a “security incident took place.” Most lacked specificity in terms of exact language that described the event more vividly. This manner made the victim aware of the incident but did not provide detailed information. Even the steps needed to remedy the intrusion were similar in nature. Four popular actions included apologies and notes of how serious data privacy is taken, the hardening of existing security controls, examination of current policies and procedures, and increased employee training. 

\section{Conclusion }
This study explored the changing cybersecurity landscape during the COVID-19 pandemic, where remote work became common. The study's findings have important lessons for organizations, governments, and institutions worried about cybersecurity and remote work. The study provides information that goes beyond Montana and can be used to shape global cybersecurity strategies.
By closely examining the dataset, this study aims to draw conclusions on trends that will prove useful for IT professionals in enhancing cybersecurity practices. The analysis encompasses breaches of varying sizes, complexities, and service interruptions. Our study shows that analyzing data can help us learn from the past and prepare for an uncertain future. Therefore, organizations can use the data to find weaknesses in their security and improve their defenses in the ever-changing world of cybersecurity. This could mean implementing better security measures or investing in advanced monitoring.
The study suggests that organizations should take care of people affected by data breaches, showing how important consumer protection is. Notably, a significant percentage of breached organizations offered Identity/Credit Monitoring services to victims. Moreover, an outside forensics firm or third-party security professionals often assisted in breach discovery and remediation planning. The notification letters sent to customers displayed similarities in language, structure, and steps taken to address the breach. While each breach was unique, the overall wording and actions were consistent. However, the notification letters lacked detailed information about the incidents and the specific actions required for remediation.
The study also encourages more research, covering different places, to get a bigger view of cybersecurity worldwide. As the global economy adapts to a post-pandemic world, where safety and being ready for anything are crucial, this study provides guidance for organizations that want to protect their data and keep their operations safe. As the trend of remote work continues to evolve beyond the pandemic, it is essential for organizations to remain vigilant and adaptable in their cybersecurity strategies. 
It is crucial to keep in mind that this study has certain limitations. One of the primary limitations is the relatively small number of cases in Montana. Although the data provides valuable insights into cybersecurity in the state, further research is required to understand how it compares to the global scenario. Future research should focus on evaluating the long-term effects of remote work on cybersecurity, exploring emerging threats and technologies, and identifying effective measures to foster a culture of cybersecurity in remote work settings. By building upon the existing knowledge base, organizations can navigate the challenges of remote work while safeguarding their critical assets and maintaining the trust of their stakeholders.

\bibliographystyle{ieeetr}
\bibliography{references}

\end{document}